\documentclass{svmult}

\usepackage{graphicx}
\usepackage{graphicx}
\usepackage{amsmath}
\usepackage{amssymb}
\usepackage{dsfont}
\usepackage{url}
\usepackage{cite} % damit aufeinanderfolgende Zitate zusammengefasst werden

\graphicspath{{.}{../}{./eps/}}

\begin{document}

\title*{Comparative numerical study of Anderson localisation in disordered
       electron systems}
\titlerunning{Anderson localisation in disordered
       electron systems}
\author{Gerald Schubert\inst{1} \and Alexander Wei{\ss}e\inst{2} \and
Gerhard Wellein\inst{3} \and Holger Fehske\inst{1}}
\authorrunning{G. Schubert, A. Wei{\ss}e, G.~Wellein, H.~Fehske}
\institute{Ernst-Moritz-Arndt-Universit\"at
    Greifswald, Institut f\"ur Physik, Domstr.\ 10a, D-17489
    Greifs\-wald, Ger\-many\and
   School of Physics, The University of New South Wales,
    Sydney, NSW 2052, Australia\and
   Regionales Rechenzentrum Erlangen (RRZE),
    Martensstra{\ss}e 1, D-91058 Erlangen, Germany}
\maketitle

\abstract{
  Taking into account that a proper description of disordered systems
  should focus on distribution functions, the authors develop a
  powerful numerical scheme for the determination of the probability
  distribution of the local density of states (LDOS), which is based
  on a Chebyshev expansion with kernel polynomial refinement and
  allows the study of large finite clusters (up to $100^3$). For the
  three-dimensional Anderson model it is demonstrated that the
  distribution of the LDOS shows a significant change at the disorder
  induced delocalisation-localisation transition. Consequently, the
  so-called typical density of states, defined as the geometric mean
  of the LDOS, emerges as a natural order parameter. The calculation
  of the phase diagram of the Anderson model proves the efficiency and
  reliability of the proposed approach in comparison to other
  localisation criteria, which rely, e.g., on the decay of the
  wavefunction or the inverse participation number.
}

%\pacs{72.15.Rn,  71.23.An, 71.30.+h, 05.60.Gg, 71.55.Jv}

%  72.15.Rn Localization effects (Anderson or weak localization)
%  71.23.An Theories and models; localized states
%  05.60.Gg Quantum transport
%  71.30.+h Metal-insulator transitions and other electronic transitions
%  71.55.Jv Disordered structures; amorphous and glassy solids

%%%%%%%%%%%%%%%%%%%%%%%%%%%%%%%%%%%%%%%%%%%%%%%%%%%%%%%%%%%%%%%%%%%%%%%%%
%                                                                       %
%   I N T R O                                                           %   
%                                                                       % 
%%%%%%%%%%%%%%%%%%%%%%%%%%%%%%%%%%%%%%%%%%%%%%%%%%%%%%%%%%%%%%%%%%%%%%%%%
\section{Introduction}
The localisation of quantum particles in disordered systems is one of
the most intensively studied problems in condensed matter
physics~\cite{Th74,We76,LR85,VW92,KM93_2}. In real systems disorder can
arise for a number reasons. We may think of randomly distributed
impurities, vacancies or dislocations in an otherwise ideal crystal,
of random arrangements of electronic or nuclear spins, etc.  While the
disorder appears in many forms that are sometimes difficult to
characterise theoretically, the randomness in the model introduced and
discussed by Anderson is simple but sufficient to capture the basic
features of the disorder-induced metal insulator
transition~\cite{An58}. The Anderson Hamiltonian,
\begin{equation}\label{And_mod}
  {H} =  - t \sum\limits_{\langle ij \rangle} 
  \bigl[{c}_i^{\dag} {c}_j^{} + \text{H.c.}\bigr]
  + \sum\limits_{j=1}^{N} \epsilon_j {c}_j^{\dag} {c}_j^{} \,,
\end{equation}
describes noninteracting electrons moving on a lattice with random
on-site potentials (compositional disorder). The operators
${c}_j^{\dag}$ (${c}_j^{}$) create (annihilate) an electron in
a Wannier state centred at site $j$, and the local potentials
$\epsilon_j$ are assumed to be independent, uniformly distributed random
variables,
\begin{equation}\label{Kasten}
   p( \epsilon_j) = \frac{1}{W} 
   \;\theta\left(\frac{W}{2}-|\epsilon_j|\right)\,.
\end{equation}
The parameter $W$ is a measure for the strength of disorder and is
usually given in units of the nearest neighbour hopping matrix element
$t$. Throughout this work we consider d-dimensional  
hyper-cubic lattices with $N=L^d$ sites, impose periodic boundary 
conditions (PBC), and set the lattice spacing equal to unity.

The spectral properties of the Anderson model~\eqref{And_mod} have
been carefully analysed (see, e.g., Ref.~\cite{FMSS85}).  For
sufficiently large disorder or near the band tails, the spectrum
consists exclusively of discrete eigenvalues, and the corresponding
eigenfunctions are exponentially localised. Since localised electrons
do not contribute to the transport of charge or energy, the energy
that separates localised and extended eigenstates is called the
mobility edge. For any finite disorder $W>0$, on a one-dimensional
(1d) lattice, all eigenstates of~\eqref{And_mod} are
localised~\cite{Bo63,MT61}. This is believed to hold also in 2d, where
the existence of a transition from localised to delocalised states at
finite $W$ would contradict the one parameter scaling
theory~\cite{AALR79,Ja98}.

In spite of the progress made over the last decades, the Anderson
metal-insulator transition is still not completely understood.  There
are several reasons why the existing theories remain unsatisfactory.
Especially when electron-electron or electron-phonon interactions
come into play, the very successful one-parameter scaling approach
might be problematic, because close to the localisation transition the
energy scales associated with both disorder and interactions are
comparable to the Fermi energy~\cite{DPN03}. On the other hand, the
numerical study of the localisation-delocalisation transition is
demanding, since the involved length scales can become extraordinary
large, in particular near the critical point. Obviously, methods that
are based on a full diagonalisation of the Hamiltonian and on the
study of the one-particle eigenstates are restricted to rather small
systems. Examples are the calculation of the localisation 
length from the decay of the electronic wavefunction or the evaluation of the
inverse participation number. In
addition, one-particle eigenstates are not defined for interacting
systems. Hence, none of these criteria can easily be generalised to
interacting disordered systems. To overcome these difficulties is
perhaps the most challenging issue of current research on disordered
materials.

Motivated by this situation, this contribution provides a (quasi
approximation free) numerical analysis of the recently revived
``local order parameter'' approach to the Anderson
transition, which, within the framework of the statistical dynamical
mean field approximation, has been successfully applied also to
correlated electron (phonon) systems~\cite{DPN03,BF02}.
Adopting a local point of view and focusing on the distribution of the
physically interesting quantities, the method follows the original
route to the localisation problem established by 
Abou-Chacra et~al.~\cite{AAT73}. In
particular, we demonstrate that for Anderson type models the
distribution of the local density of states (LDOS) can be determined
very easily by the kernel polynomial method (KPM)~\cite{SRVK96},
a refined Chebyshev expansion technique.  Based on the distribution
of the LDOS, localised states are distinguished from extended states
by a vanishing geometrical average, which is usually called the
``typical DOS''. In addition, it turns out that this quantity
characterises the disorder-induced metal-insulator transition also in
more complex systems~\cite{DPN03,BF02,BHV04,SWF04,SWF04b,ASWBF04}.

To examine the efficiency and
accuracy of the proposed LDOS-KPM approach, 
we carry out a comparative numerical study of the
localisation-delocalisation transition. In view of the wealth of known
results, the 3d Anderson model seems to be best suited for this kind
of investigation. The results we obtain for the mobility edge from
different methods allow for a detailed understanding of the typical
DOS concept and open the road towards an application to more complex
situations.

%%%%%%%%%%%%%%%%%%%%%%%%%%%%%%%%%%%%%%%%%%%%%%%%%%%%%%%%%%%%%%%%%%%%%%%%%
%                                                                       %
%   N U M E R I C A L   I N V E S T I G A T I O N   A N D E R S O N     %   
%                                                                       % 
%%%%%%%%%%%%%%%%%%%%%%%%%%%%%%%%%%%%%%%%%%%%%%%%%%%%%%%%%%%%%%%%%%%%%%%%%

\section{Anderson transition}\label{Num}
As the Anderson transition is expected only for $d>2$, in this chapter
we focus on the 3d case, for which the lack of successful analytical
approaches necessitates a numerical treatment. In contrast to the
widely used numerical transfer matrix method~\cite{MK83,PS81,PRS03},
which describes the 3d system as a quasi-1d system of finite cross
section, below we stress the bulk properties of the system and
consider cubic clusters which extend equally in all spatial
directions. Due to the large length scales that emerge in the critical
region, it is generally a difficult task to interpret the results of
such finite cluster calculations.

As a kind of benchmarking, we review and compare established localisation 
criteria, namely, the localisation length
(Sec.~\ref{DW}) and the inverse participation number (Sec.~\ref{pr}).
Both can be extracted from the one-particle wavefunctions, which,
however, requires the complete numerical diagonalisation of the
Hamiltonian~\eqref{And_mod}. In Sec.~\ref{LD}, 
we present the new approach
that is based on the distribution of the LDOS.  Since the
calculation of the LDOS via the KPM
requires only sparse matrix vector multiplications, this technique
scales linearly with the number of lattice sites and 
permits the study of
significantly larger systems.

%%%%%%%%%%%%%%%%%%%%%%%%%%%%%%%%%%%%%%%%%%%%%%%%%%%%%%%%%%%%%%%%%%%%%%%%%
%              D E C A Y   O F   W A V E F U N C T I O N                %   
%%%%%%%%%%%%%%%%%%%%%%%%%%%%%%%%%%%%%%%%%%%%%%%%%%%%%%%%%%%%%%%%%%%%%%%%%

\subsection{Decay of the wavefunction}
\label{DW}

\begin{figure}
  \begin{minipage}{0.67\linewidth}
    \includegraphics[width=0.95\linewidth]{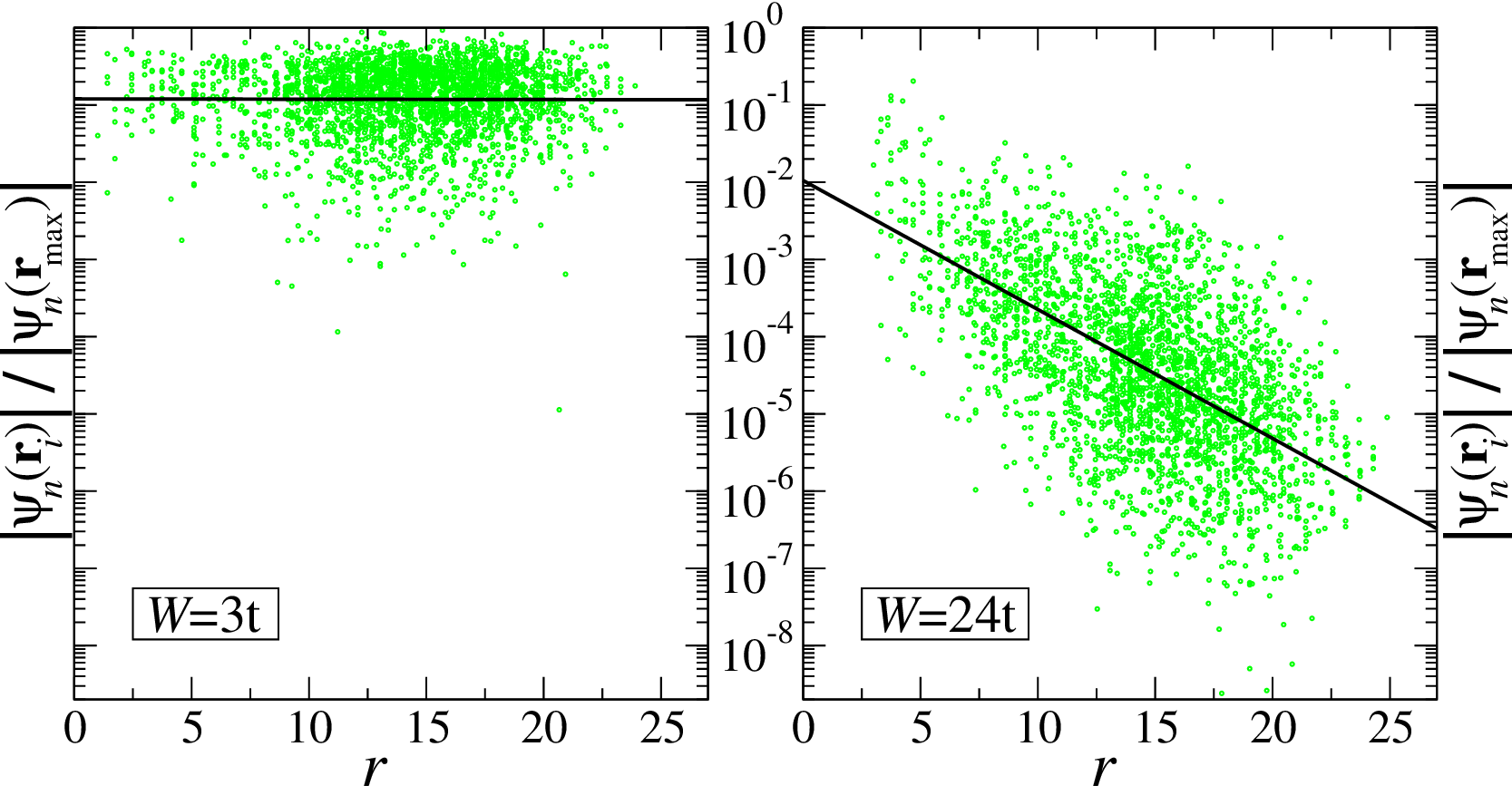}
  \end{minipage}
  \begin{minipage}{0.32\linewidth}
    {\bf Fig.1} Decay of an electronic wavefunction 
     $\psi_n$ in the band centre as a function of the
     distance $r= |{\bf r}_i - {\bf r}_{\text{max}}|$ to the site with
     maximum amplitude, ${\bf r}_{\text{max}}$. Results are given
     each for one fixed energy and realization of disorder, $N=30^3$, PBC.
  \end{minipage}
  \label{Schrot}
\end{figure}
The most obvious but costly way to access the localisation properties
of an electronic wavefunction is the direct calculation of the localisation
length $\lambda$, which is infinite for extended states and finite
otherwise. For localised states, the envelope of the wavefunction
decays exponentially from some point ${\bf r}_{\text{max}}$ in the
crystal.
\begin{equation}\label{expdecay}
  |\psi_n({\bf r}_i)| \sim f({\bf r}_i) 
  \exp\left(-\frac{|{\bf r}_i - {\bf r}_{\text{max}}|}{\lambda}\right)\,.
\end{equation}
The random function $f({\bf r}_i)$ describes the statistical
fluctuations of the amplitudes $\psi_n({\bf r}_i)$ of the
eigenfunction $\psi_n$ at energy $E_n$. Given $\psi_n$,
the localisation length $\lambda(E_n)$ is obtained
by locating the site of maximum amplitude, ${\bf r}_{\text{max}}$, and
fitting Eq.~\eqref{expdecay} to the data. In contrast to the case of
weak disorder, where the amplitude is essentially independent of the
distance from ${\bf r}_{\text{max}}$, at higher values of $W$ a clear
exponential decay is observed (see Fig.~1). Note, that
besides the direct fit with equal weight for the amplitudes of all
sites, $\lambda$ can also be determined using the method of
asymptotic slope~\cite{LT74}. Here the data is first averaged within
shells of fixed distance from ${\bf r}_{\text{max}}$ and fitted
thereafter. However, using this second approach the detection of the
Anderson transition is not as robust and more sensitive to the
fluctuations of the data. We therefore refrain from considering
corresponding results.

%%%%%%%%%%%%%%%%%%%%%%%%%%%%%%%%%%%%%%%%%%%%%%%%%%%%%%%%%%%%%%%%%%%%%%%%%
%         I N V E R S E   P A R T I C I P A T I O N   N U M B E R       %
%%%%%%%%%%%%%%%%%%%%%%%%%%%%%%%%%%%%%%%%%%%%%%%%%%%%%%%%%%%%%%%%%%%%%%%%%
\subsection{Inverse participation number}\label{pr}

\begin{figure}
  \begin{minipage}{0.63\linewidth}
    \includegraphics[width=0.95\linewidth]{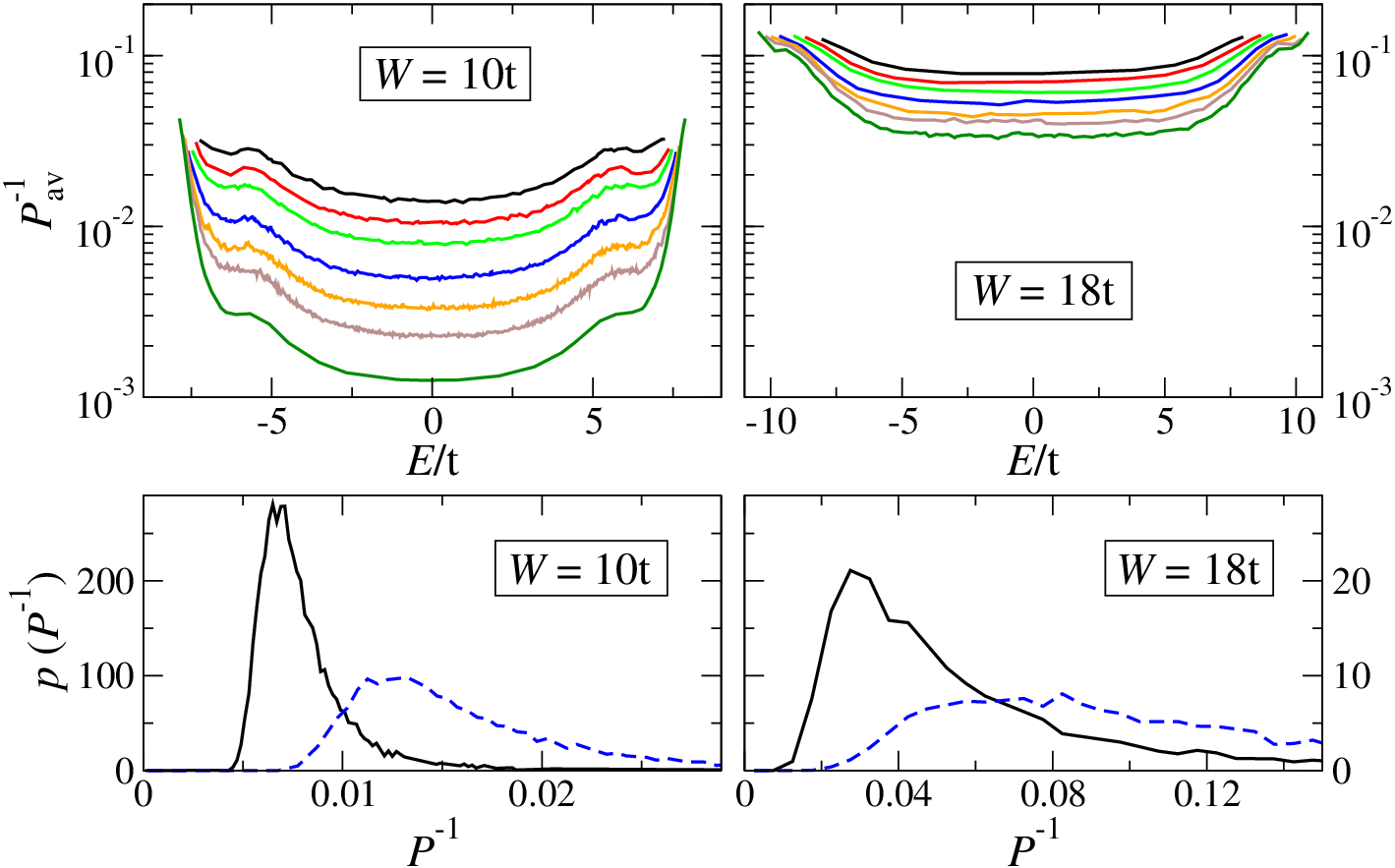}
  \end{minipage}
  \begin{minipage}{0.36\linewidth}
    {\bf Fig.2} Upper part: Averaged inverse participation number 
     $P^{-1}_{\text{av}}$ of $1000$ systems with $L^3$ sites and PBC 
     (top to bottom: $L=8, 9, 10, 12, 14, 16, 20$). Lower part: 
     Probability
     density of $P^{-1}$ for the $10^3$ system in the band centre
     (solid line) and near the band edges (dashed line). Note the
     different scales in the lower panels.
  \end{minipage}
  \label{IPN}
\end{figure}
Yet another quantity that measures the Anderson transition is the
inverse participation number~\cite{We80},
\begin{equation}
  P^{-1}(E_n) = \sum\limits_{i=1}^{N} |\psi_n({\bf r}_i)|^4\,,
\end{equation}
which is proportional to the inverse number of sites that contribute
to a given one-particle wavefunction $\psi_n$.  For
delocalised states we find $P^{-1}\sim 1/N$, which vanishes in the
thermodynamic limit. Localised states, on the other hand,
approximately extend over a finite volume $N_0$, yielding $P^{-1} \sim
1/N_0$ independently of the system size $N$. In Fig.~2 this
behaviour is demonstrated for small and large disorder $W$.  While in
the localised case ($W=18t$) $P^{-1}$ is almost independent of $L$,
apparently it decreases with $L$ for extended states ($W=10t$). Apart 
from the different
scaling, the distribution of $P^{-1}$ changes at the transition, being
sharply peaked for extended states and rather broad for localised ones
(lower part of Fig.~2).
\begin{figure}
  \begin{minipage}{0.65\linewidth}
    \includegraphics[width=0.95\linewidth]{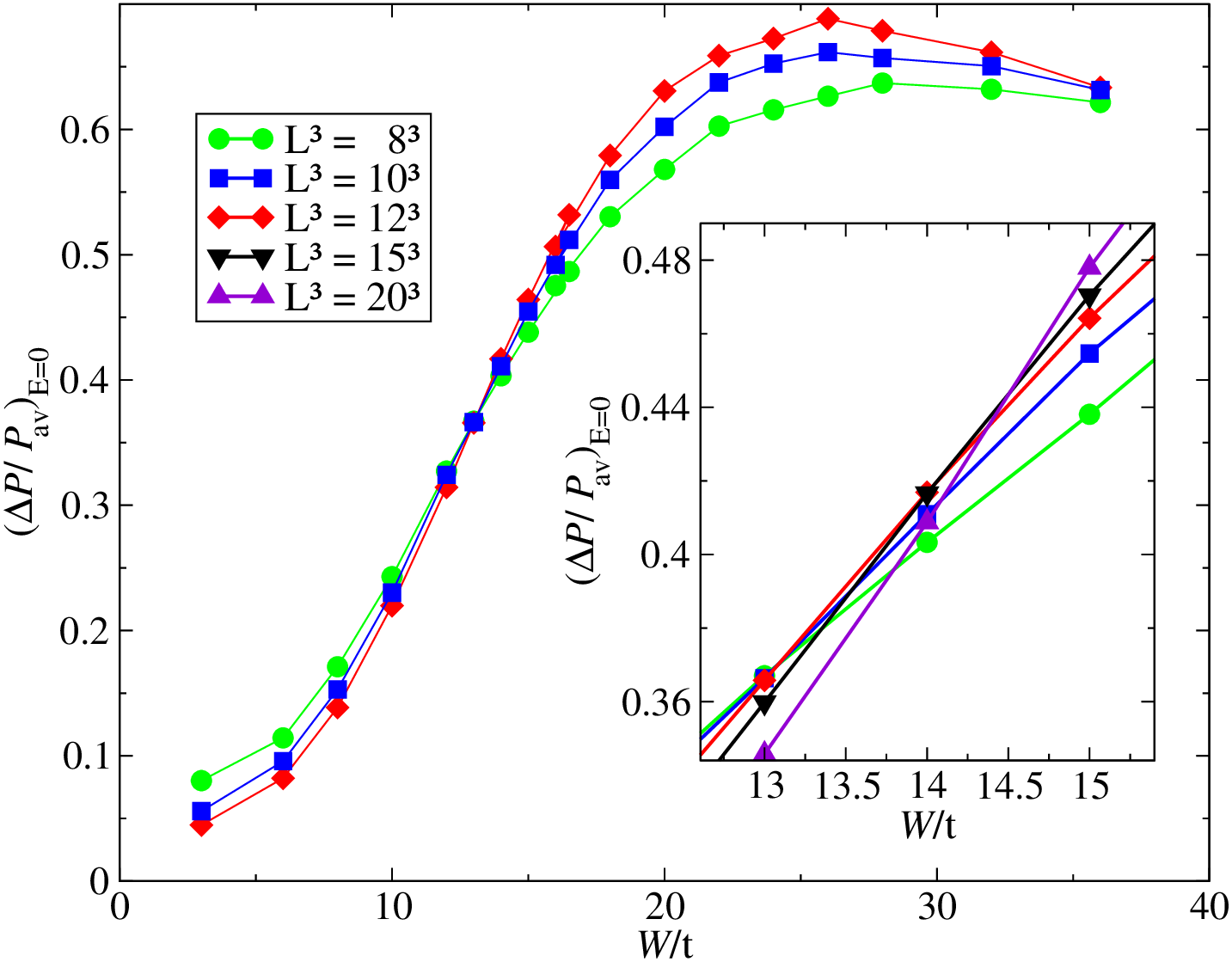}
  \end{minipage}
  \begin{minipage}{0.34\linewidth}
    {\bf Fig.3} Normalised standard deviation of the
    participation number
    in the band centre as a function of disorder for
    different system sizes using PBC.  The obtained results were
    averaged over $1000$ realizations of disorder.
  \end{minipage}
  \label{Spanier}
\end{figure}

Based on the distribution of the participation number $P$ recently an
alternative numerical approach for monitoring the Anderson transition
was proposed~\cite{MMD03}. In analogy to results for a certain class of
power-law random banded matrices, which indicate the scale invariance
of the distribution of $P$ at the Anderson transition, Malyshev et~al.~
\cite{MMD03} suggest to detect the transition by studying the
ratio of the standard deviation of $P$, $\Delta P$, to the mean
participation number $P_{\text{av}}$, which should be independent 
of the system size at the critical disorder.

So far this approach has only been tested for a one-dimensional model
with diagonal disorder and power-law long-range hopping~\cite{MMD03},
which shows a transition at the band edge and can thus be tackled with
the Lanczos method.  In Fig.~3 we present first data for
the band centre of the standard Anderson model. While for small
disorder the ratio $\Delta P/P_{\text{av}}$ decreases with increasing
system size, at large disorder the opposite happens.  The intersection
of the curves is not completely independent of the system size (see
inset of Fig.~3), and a precise determination of the
transition requires some finite size scaling of the data.  
Performing a finite-size scaling~\cite{MMD03} 
our data is consistent with a
critical disorder strength of $W_c\approx 16.1\pm 0.8$ in the
thermodynamic limit.

%%%%%%%%%%%%%%%%%%%%%%%%%%%%%%%%%%%%%%%%%%%%%%%%%%%%%%%%%%%%%%%%%%%%%%%%%
%         L O C A L   D E N S I T Y   O F   S T A T E S                 %
%%%%%%%%%%%%%%%%%%%%%%%%%%%%%%%%%%%%%%%%%%%%%%%%%%%%%%%%%%%%%%%%%%%%%%%%%

\subsection{Local density of states}\label{LD}

\begin{figure}%[ht]
  \begin{minipage}{0.67\linewidth}
    \includegraphics[width=0.95\linewidth]{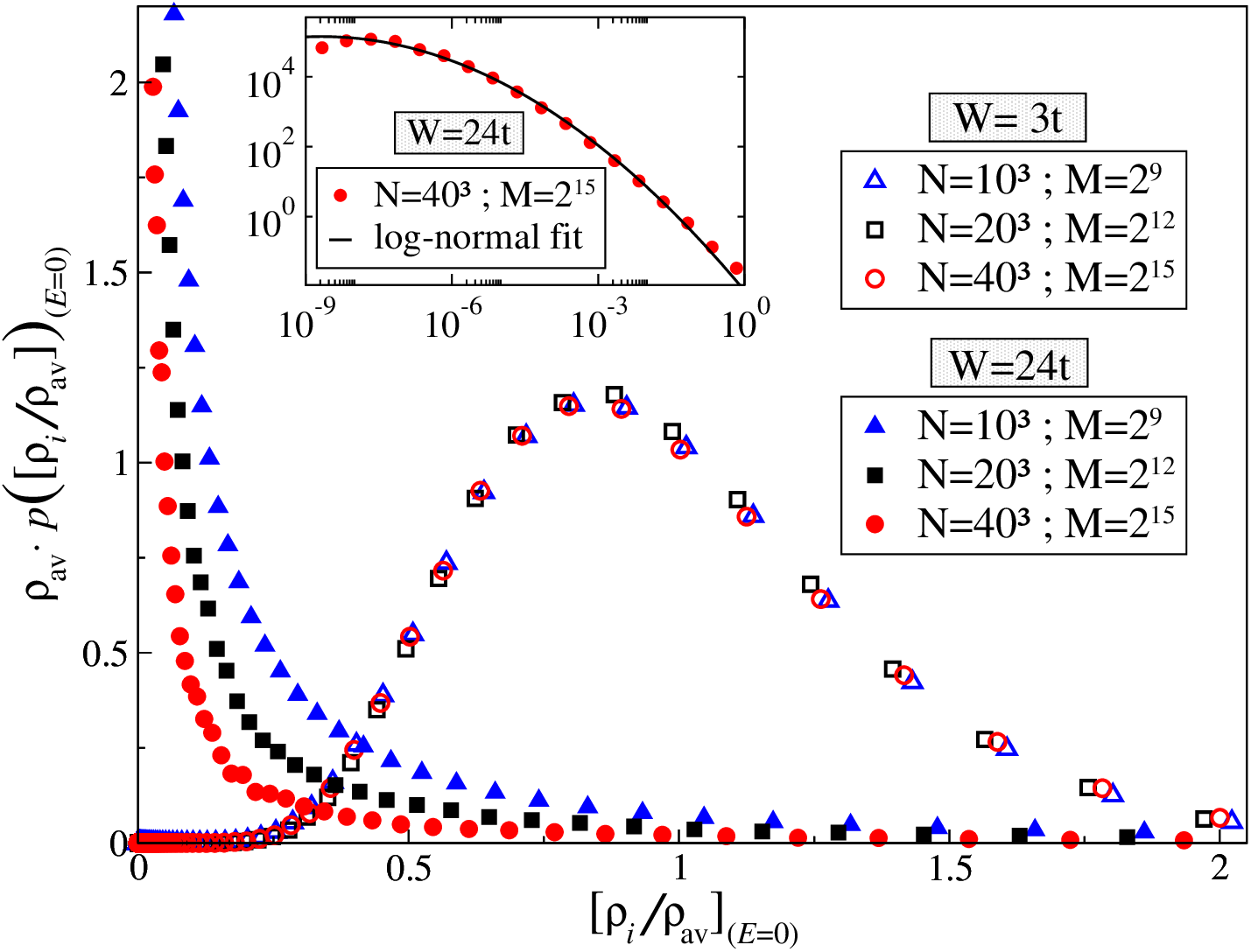}
    \label{Vfkt_LDOS_2}
  \end{minipage}
  \begin{minipage}{0.32\linewidth}
    {\bf Fig.4} General shape and finite size 
       scaling of the LDOS distribution $p(\rho_i/\rho_{\text{av}})$.
       Keeping the ratio
       $N/M=1.95$ fixed for $N=10^3,\, 20^3,\ 40^3$
       and $K_r\times K_s=10^4\times 100,\, 100\times100,\,32\times32$
       respectively, we calculated histograms for 
       $E \in [-0.1t,0.1t]$.  Inset:
       Double logarithmic plot of $p(\rho_i/\rho_{\text{av}})$
       for the localised
       case
       together with a
       log-normal fit to the data.
  \end{minipage}
\end{figure}

Already in his seminal paper~\cite{An58} Anderson pointed out that in
order to describe the transition from delocalised to localised states
it is very instructive to discuss the distribution of local quantities
of interest, such as the escape rate or recurrence probability from or
to a given site. Another suitable quantity that becomes critical at
the Anderson transition is the LDOS~\cite{HT94,DPN03},
\begin{equation} \label{LDOS}
  \rho_i(E) = \sum\limits_{n=1}^{N}
  | \psi_n ({\bf r}_i)|^2 \delta(E-E_n)\,,
\end{equation}
which for a given energy directly measures the local amplitude of the
wavefunction at site ${\bf r}_i$. So far the LDOS has been considered
mainly within analytical approaches or by the Lanczos recursion
method~\cite{HT94}. Typically the latter suffers from severe stability
problems at high expansion order and conclusive results for the
Anderson transition are difficult to obtain~\cite{AH02}. Fortunately
the KPM technique~\cite{SRVK96} described in
Appendix~\ref{Anhang_KPM} is a very efficient way to circumvent these
difficulties and allows the calculation of high-resolution LDOS data
for very large systems. In a nutshell, within this approach the
function of interest is expanded in a finite series of Chebyshev
polynomials. To weaken the effects of the truncation and ensure
properties such as positivity and normalisation, the function is
convoluted with an appropriate integral kernel. The resolution of the
method is inversely proportional to the order of the expansion $M$
(the number of so-called Chebyshev moments).

Adopting Anderson's original point of view that a proper description
of disordered systems should focus on distribution functions, we
calculated $\rho_i(E)$ for a large number of samples, $K_r$, and
sites, $K_s$, and studied its statistical properties. In
Figure~4 we show the resulting distribution of
$\rho_i(E=0)$, normalised by its mean value $\rho_{\text{av}}$, for
two characteristic values of disorder. 
As $\rho_{\text{av}}$ is a function of disorder, this normalisation 
ensures $\langle\rho_i/\rho_{\text{av}}\rangle=1$ independent of 
$W$, allowing thus an appropriate comparison. In the
delocalised phase, $W=3t$, the distribution is rather symmetric and
peaked close to its mean value. Note that increasing the system
size and the expansion order, such that the ratio of mean
level spacing and KPM resolution is fixed, does not change the
distribution. This is in strong contrast to the localised phase, e.g.,
$W=24t$, where the distribution of $\rho_i(E)$ is extremely asymmetric.
Although most of the weight is now concentrated close to
zero, the distribution extends to very large values of $\rho_i$,
causing the mean value to be much larger than the most probable
value. In addition, a similar finite size scaling increases the
asymmetry and underlines the singular behaviour expected in the
thermodynamic limit and at infinite resolution. Note also, that the
distribution of the LDOS is well approximated by a log-normal
distribution~\cite{MS83},
\begin{equation}
  p(x) = \frac{1}{\sqrt{2\pi\sigma^2}} \frac{1}{x}
  \exp\left(-\frac{\left(\ln\left(x/x_0\right)\right)^2}
    {2\sigma^2}\right)\,,
\end{equation}
as illustrated in the inset of Fig.~4.
\begin{figure}
  \begin{minipage}{0.75\linewidth}
      \includegraphics[width=0.95\linewidth]{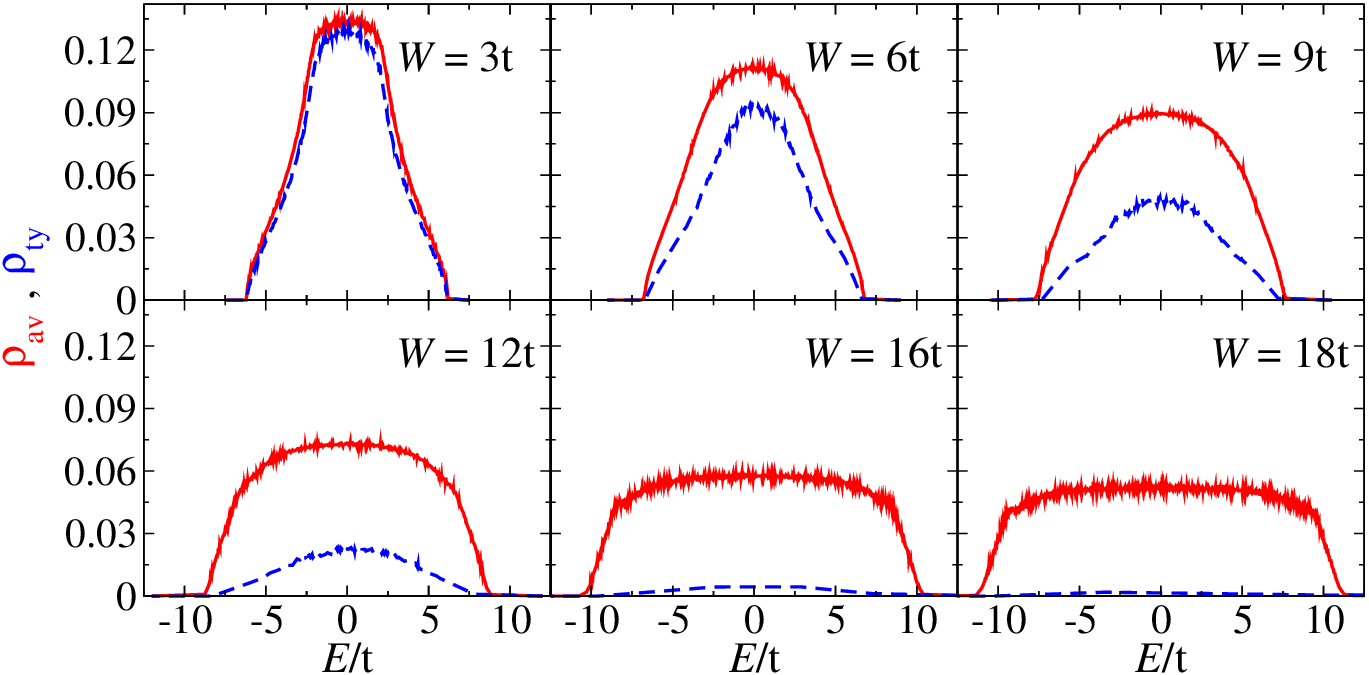}
  \end{minipage}
  \begin{minipage}{0.24\linewidth}
    {\bf Fig.5} Average (solid line) and typical (dashed line) 
     DOS for a $50^3$ lattice with
     PBC. $K_s\times K_r = 32\times 32$, $M=8192$.
  \end{minipage}
  \label{av_ty_DOS}
\end{figure}

Of course, the study of entire distributions is a bit inconvenient,
and for practical calculations, instead, we will prefer an appropriate
statistics that uniquely characterises the distribution. The above
findings suggest, that such a statistics is given by the arithmetic
and geometric averages of $\rho_i(E)$,
\begin{align}
  \rho_{\text{av}}(E) & = \frac{1}{K_r K_s} 
  \sum\limits_{k=1}^{K_r} \sum\limits_{i=1}^{K_s}\rho_i(E)\,,\label{rhoav}\\
   \rho_{\text{ty}}(E) & = \exp \left(\frac{1}{K_r K_s} \smash{
   \sum\limits_{k=1}^{K_r}\sum\limits_{i=1}^{K_s}
   \ln\bigl(\rho_i(E)\bigr)} \right) \,. \label{rhoty}
\end{align}
On the one hand, the arithmetic mean for large enough
  $K_r$ and $K_s$ converges to the standard density of states $\rho(E)
  = \sum_{n=1}^{N} \delta(E-E_n)$, which is not critical at the
  Anderson transition. The geometric mean, on the other hand,
  represents the typical value of the distribution, which, as shown
  above, is finite in the delocalised phase, but goes to zero in the
  localised phase. As can be seen from Fig.~5,
  $\rho_{\text{av}}(E)$ and $\rho_{\text{ty}}(E)$ are almost equal for
  extended states, whereas for localised states $\rho_{\text{ty}}(E)$
  vanishes and $\rho_{\text{av}}(E)$ remains finite.  This implies,
  that the ratio of these two quantities, the normalised typical
  density of states
\begin{equation}
  R(E) := \frac{\rho_{\text{ty}}(E)}{\rho_{\text{av}}(E)}\,,
\end{equation}
can serve as an order parameter for the Anderson transition with $R>0$
for extended states and $R=0$ for localised ones.
\begin{figure}%[ht]
  \begin{minipage}{0.75\linewidth}
    %\centering 
    \includegraphics[width=0.95\linewidth]{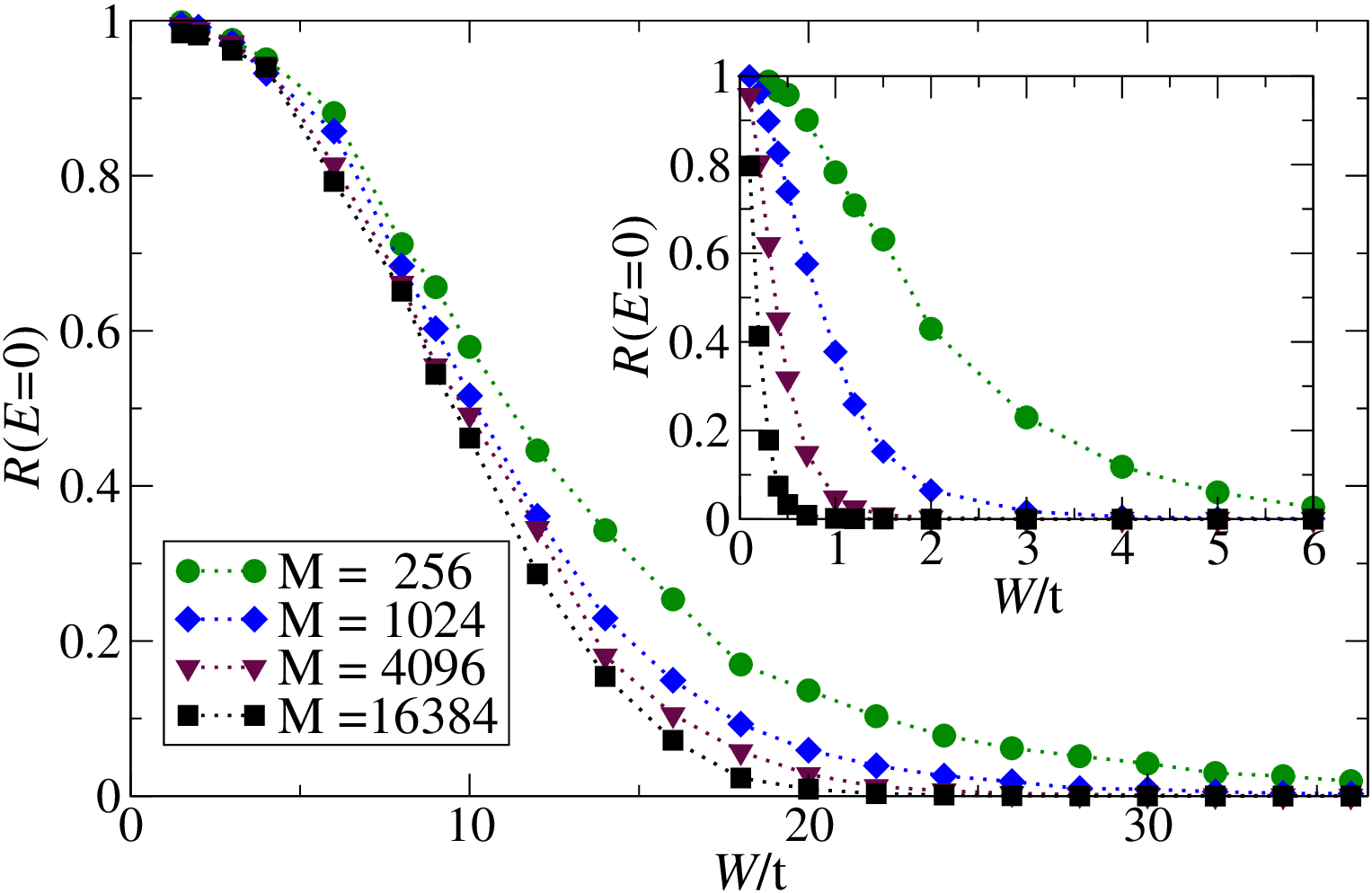}
  \end{minipage}
  \begin{minipage}{0.24\linewidth}
    {\bf Fig.6} Normalised typical DOS as a function of 
     disorder calculated with increasing expansion order on a
     $50^3$ lattice. The inset shows the corresponding behaviour of the
     1d system with $125000$ sites. $K_s\times K_r = 32\times 32$.
  \end{minipage}
  \label{tydos_M}
\end{figure}
 As for most
numerical calculations the transition is slightly washed out by the
finite size of the considered cluster, and by the KPM resolution. 
However, for large clusters and increasing $M$ a plot of the
ratio $R$ versus disorder strength $W$ (see Fig.~6) allows
for a reliable determination of the critical disorder $W_c$, and,
e.g., in the band centre we obtain $W_c(E=0) \simeq 16.5t$ in
accordance with other numerical results for the 3d Anderson
model~\cite{BSK87,KM93_2,GS95}. The quality of this criterion is
underlined also by our data for a 1d system shown in the inset of
Fig.~6. Here, as mentioned above, arbitrarily small
disorder leads to localisation of the entire spectrum. Clearly, in our
approach this is reflected by a typical DOS that vanishes for large $M$.

%
%%%%%%%%%%%%%%%%%%%%%%%%%%%%%%%%%%%%%%%%%%%%%%%%%%%%%%%%%%%%%%%%%%%%%%%%%
%                          C O M P A R I S O N                          %
%%%%%%%%%%%%%%%%%%%%%%%%%%%%%%%%%%%%%%%%%%%%%%%%%%%%%%%%%%%%%%%%%%%%%%%%%
%
%
\subsection{Comparison of the different methods}

 \begin{figure}[b]
   \begin{center}
      \includegraphics[width=0.85\linewidth]{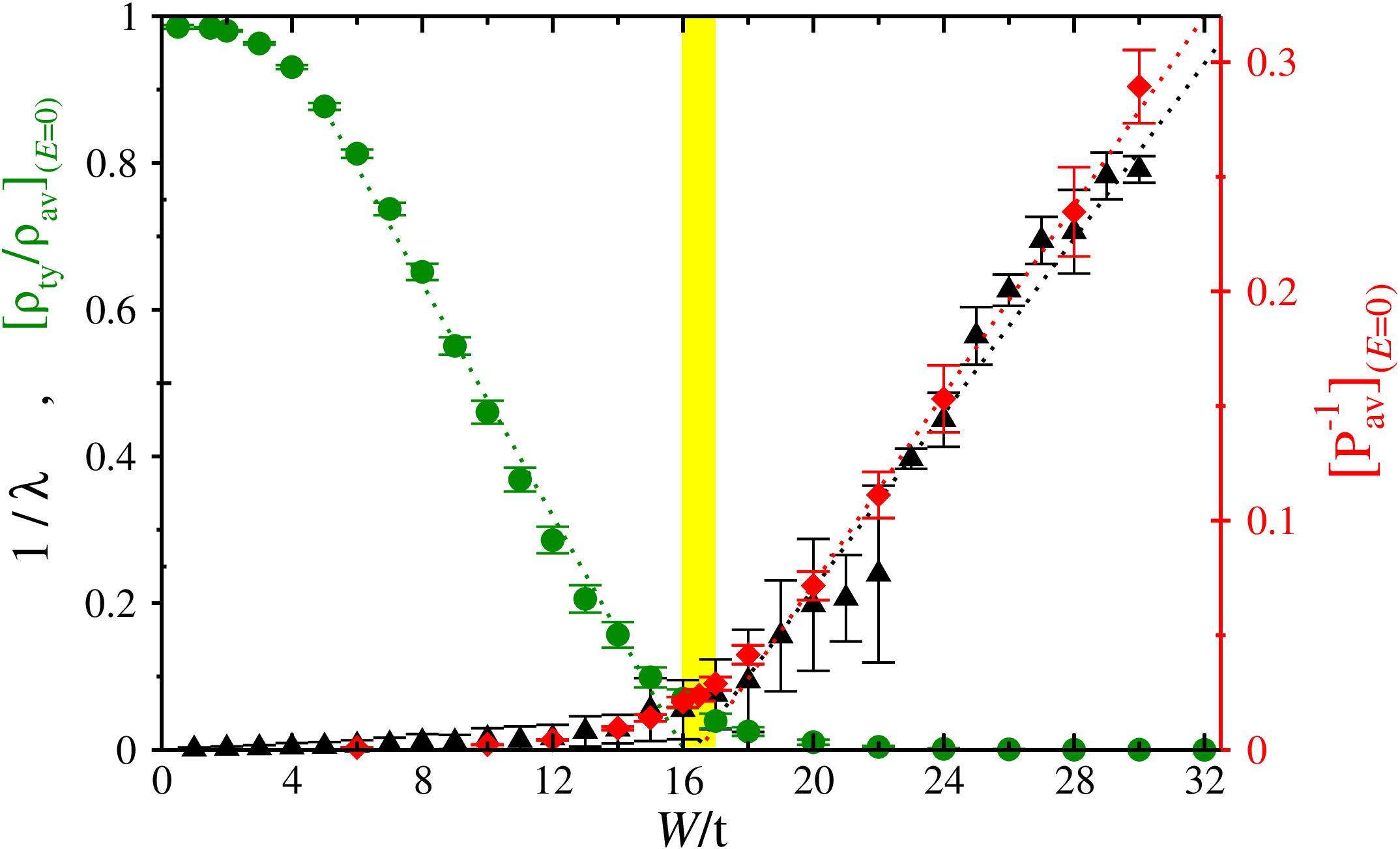}
   \end{center}	
     {\bf Fig.7} Comparison of the critical values of 
     disorder, $W_c$, obtained by the methods outlined in Secs.~2.1-2.3.
     Decay of the wavefunction (triangles): $N = 20^3$ and $30^3$, 
     $E_n \in [-0.01,0.01]$, $K_r=10$.
     Average inverse participation number (diamonds): $N=16^3$, 
     $E_n \in [-0.1,0.1]$, $K_r = 100$.
     Normalised typical DOS (circles): $N=50^3$, $M=16384$,  
     $K_r\times K_s=32\times32$.
   \label{Vgl_Dresden}
 \end{figure}

Comparing the value of the critical disorder obtained by the various
methods discussed in the previous sections, the two main results are
the following: (i) As can be seen from Fig.~7, 
the established criteria and methods show an uncertainty of the
critical value $W_c$ in the order of $\pm 0.5t$, which is mainly due
to the finite system sizes accessible to the numerical calculations.
Note that our data widely agrees with the results in the
literature~\cite{KBMS90,KM93_2}. 
(ii) The value $W_c\simeq 16.5t$ can be
reproduced with the same accuracy using a vanishing typical DOS as an
indicator for localisation.  An improvement of the accuracy of this
result can in principle be obtained by extending the numerical effort
(larger systems, higher resolution, high-performance computers), which
is facilitated by the straightforward parallelisability of the KPM
algorithm. On the other hand, an appropriate scaling ansatz may
improve the estimate of $W_c$ on the basis of the presented data.
 
Using the well-established value $W_c(E=0)\simeq 16.5t$ as a
calibration of the critical $R$, required to distinguish localised
from extended states for the used values of $N$ and $M$, we reproduce the
mobility edge in the energy-disorder plane~\cite{GS95,KM93_2}
using $R_c\simeq 0.05$ (see Fig.~8). We also find the
well-known reentrant behaviour near the unperturbed band
edges~\cite{BSK87,Qu01}: Varying $W$ for some fixed values of $E$
($6t<E\le 7.6t$) a region of extended states separates two regions
of localised states. The Lifshitz boundaries, shown as dashed lines,
indicate the energy range, where eigenstates are in principle allowed.
As the probability of reaching the Lifshitz boundaries is
exponentially small, we cannot expect to find states near these
boundaries for the finite ensembles considered in any numerical
calculation.

With respect to numerical resources, clearly, the methods that are
based on a complete diagonalisation of the Hamiltonian (decay
of the wavefunction or participation number) are the least
favourable ones, since their CPU requirements scale as $N^3$ and the
memory as $N^2$.  Using LAPACK~\cite{LAPACK} routines for dense
matrices on a standard PC-system diagonalisations are feasible for
systems up to $21^3$. Banded matrix routines together with the
bandwidth reduction described in Appendix~\ref{RedBB} allow to
increase this size to $30^3$.

The calculation of the LDOS via KPM is based on sparse matrix vector
multiplications, whose CPU and memory requirements scale only linearly
in $N$. Hence, systems up to $100^3$ can be easily handled with
desktop computers, and the use of high-performance environments permits
the study of even larger ensembles and systems. We conclude
that the new method substantially increases the size of numerically
accessible systems, which may lead to a more thorough understanding of
the Anderson transition.
\begin{figure}
  \begin{minipage}{0.6\linewidth}
     \includegraphics[width=0.95\linewidth]{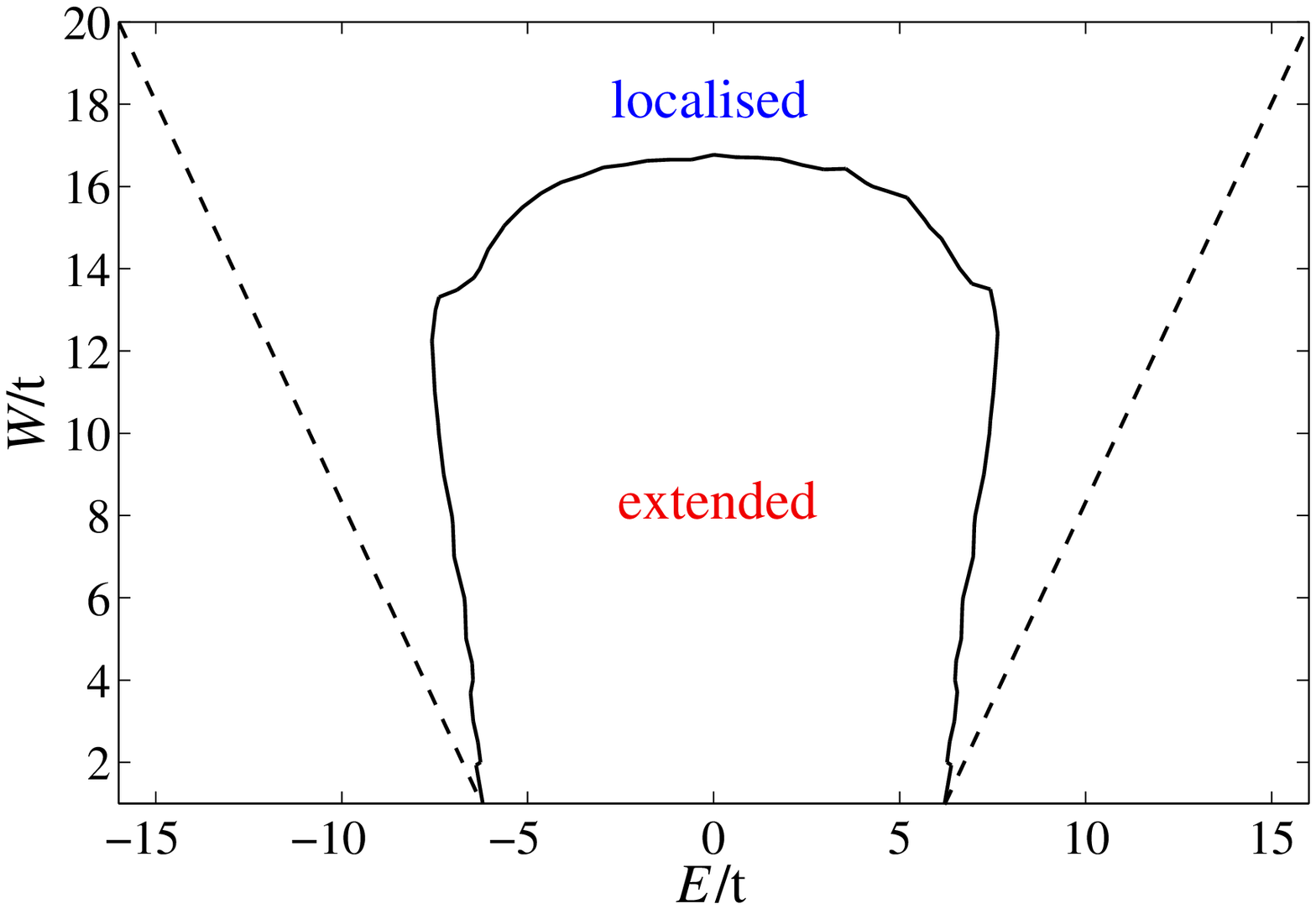}
  \end{minipage}
  \begin{minipage}{0.39\linewidth}
    {\bf Fig.8} Phase diagram of the Anderson model on a 3d cubic
     lattice. Shown are the mobility edge (solid curve) as well as the
     Lifshitz boundaries (dashed lines).
  \end{minipage}
  \label{Lifshitz}
\end{figure}

%%%%%%%%%%%%%%%%%%%%%%%%%%%%%%%%%%%%%%%%%%%%%%%%%%%%%%%%%%%%%%%%%%%%%%%%%
%                                                                       %
%                 C O N C L U S I O N S                                 %   
%                                                                       % 
%%%%%%%%%%%%%%%%%%%%%%%%%%%%%%%%%%%%%%%%%%%%%%%%%%%%%%%%%%%%%%%%%%%%%%%%%

\section{Conclusions}
With this contribution we aimed to compare well-established numerical
localisation criteria for electrons in disordered systems with a
recently proposed approach that is based on the evaluation of the
typical density of states.

Considering the 3d cubic Anderson model we proved that the local DOS
can be very efficiently calculated using a Chebyshev expansion with
kernel polynomial refinement. Given the numerically obtained
distribution of the LDOS, the corresponding typical DOS allows for the
detection of the delocalisation-localisation transition with a
precision that is comparable to results known from other methods. Like
for all numerical schemes, the method is restricted to finite systems,
and the obtained critical values are subjected to finite-size effects.
However, the low computational resources required by our approach
substantially increases the accessible system sizes.

Finally we established the use of the typical DOS as a kind of 
order parameter, which is important in view of its application to
interacting disordered systems.

\section*{}
The authors greatly acknowledge support from the Competence Network for
Technical/Scientific High-Performance Computing in Bavaria (KONWIHR).
Special thanks
go to LRZ M\"unchen, NIC J\"ulich and HLRN (Zuse-Institut Berlin) for
granting resources on their computing facilities. Discussions with
A.~Alvermann, F.X.~Bronold, S.A.~Trugman
and W.~Weller were greatly appreciated.

%%%%%%%%%%%%%%%%%%%%%%%%%%%%%%%%%%%%%%%%%%%%%%%%%%%%%%%%%%%%%%%%%%%%%%%%%
%                                                                       %
%                 A P P E N D I C E S                                   %   
%                                                                       % 
%%%%%%%%%%%%%%%%%%%%%%%%%%%%%%%%%%%%%%%%%%%%%%%%%%%%%%%%%%%%%%%%%%%%%%%%%

\appendix
  
\section{Calculation of the LDOS via the kernel polynomial method}
\label{Anhang_KPM}

At first glance, Eq.~\eqref{LDOS} suggests that the calculation
of the LDOS could require a complete diagonalisation of $ H$. It
turns out, however, that an expansion of $\rho_i$ in terms of
Chebyshev polynomials $T_n(x) = \cos(n\arccos x)$ allows for an
incredibly precise approximation. Since the Chebyshev polynomials form
an orthogonal set on the interval $[-1,1]$, prior to an expansion the
Hamiltonian $ H$ needs to be rescaled,
\begin{equation}
  {\mathfrak X} = \frac{{H}}{W/2+2dt+0.01t}\,.
\end{equation} 
Here $W/2+2dt$ reflects half the bandwidth of the Anderson
model and $0.01t$ is an additional offset that ensures numerical
stability of the expansion.
In terms of the coefficients 
\begin{equation}
 \mu_m = \int\limits_{-1}^{1} \rho_{i}(x) T_m(x) \, dx 
       = \sum\limits_{n=1}^{N}  \langle i  
         | n \rangle \langle n | T_m(x_n) | i \rangle 
       = \langle i | T_m({\mathfrak X}) | i \rangle 
    \label{mu_Sp}
\end{equation}
the approximate LDOS $\tilde{\rho}_i(x)$ reads
\begin{equation}
  \tilde{\rho}_i(x) = \frac{1}{\pi\sqrt{1-x^2}}\left( g_0\mu_0 + 
    2\sum\limits_{m=1}^{M} g_m\mu_m T_m(x)\right) \,.
\end{equation}
The factors 
\begin{equation}\label{Gibbs_Koeff}
 g_m = \frac{1}{M+1}
	 \left( (M-m+1)\cos(m\phi) + 
		\frac{\sin(m\phi)}{\tan(\phi)} \right)\, ,
\end{equation}
where $\phi = \pi/(M+1)$,
result from a convolution of the finite series with the so-called
Jackson kernel~\cite{SRVK96}, which mainly damps out the Gibbs
oscillations known from polynomial approximations (cf.
  Fig.~9).  The width of the kernel, $\Delta
x=\pi/M$, scales inversely with the order of the expansion $M$ and
defines the resolution of the method. 
\begin{figure}%[ht]
  \begin{minipage}{0.55\linewidth}
    \includegraphics[width=0.95\linewidth]{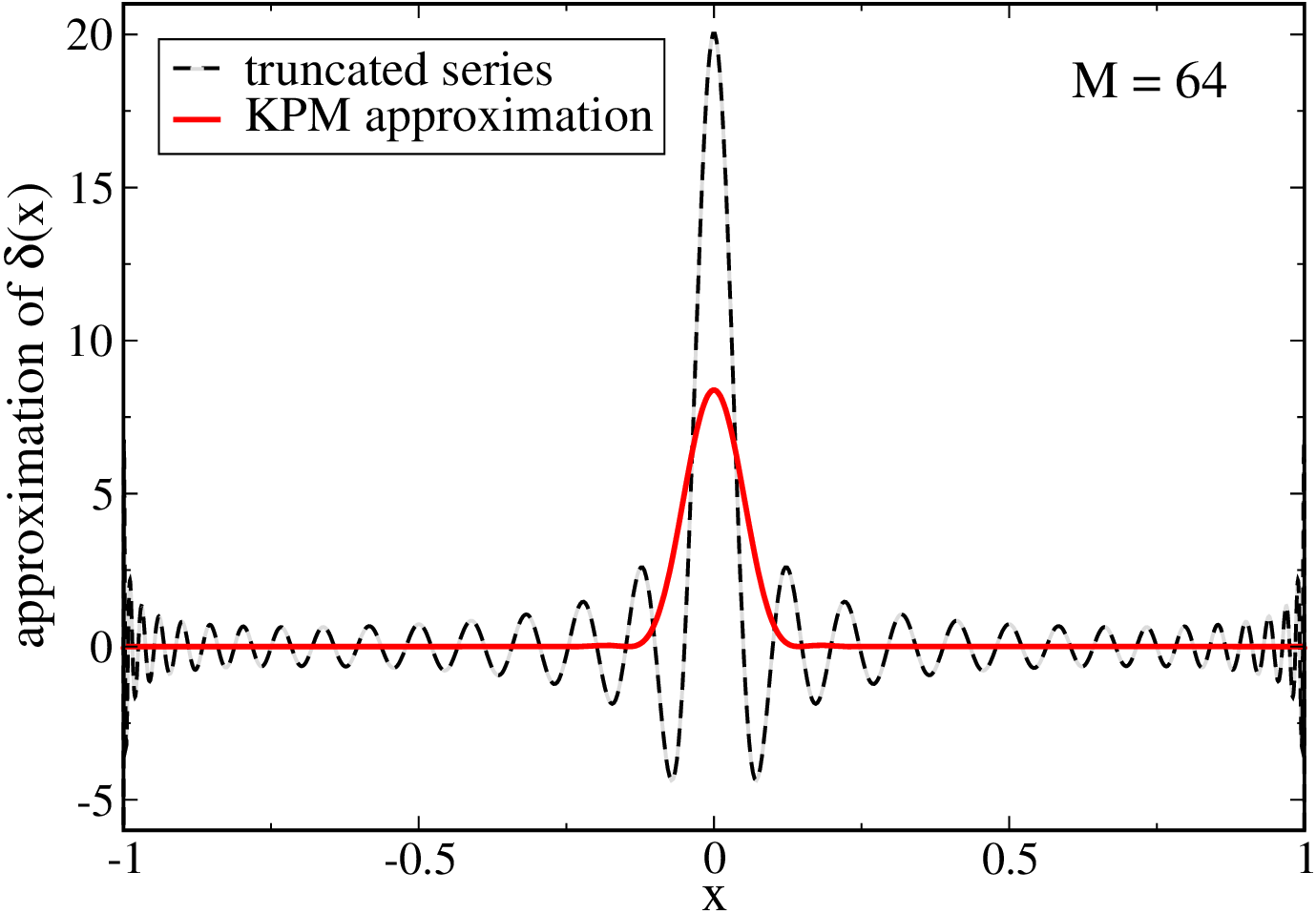}
  \end{minipage}
  \begin{minipage}{0.44\linewidth}
    {\bf Fig.9} 
    Chebyshev expansion of a $\delta$-peak: The plain truncated series 
    of order $M=64$ is a strongly oscillating curve (dashed). By convolution
    with the Jackson kernel it transforms into a strictly positive, well 
    localised peak at $x=0$ (solid), which is much closer to our usual 
    notion of $\delta(x)$.
  \end{minipage}
  \label{Jackson_Kernel}
\end{figure}

Using the recursion relations of the Chebyshev polynomials, 
\begin{equation}
  T_{m+1}(x) = 2 x T_{m}(x) - T_{m-1}(x)\,,
\end{equation}
the moments $\mu_{m}$ can be calculated iteratively.  An additional
trick allows for the generation of two moments with each matrix vector
multiplication by ${\mathfrak X}$,
\begin{equation}\label{Momcalc}
  \begin{split}
    \mu_{2m-1} & = \sum\limits_{i=1}^{N} 2
    \langle i |T_m({\mathfrak X})T_{m-1}({\mathfrak X})| i \rangle
    -  \mu_{1}\,,\\
    \mu_{2m}   & = \sum\limits_{i=1}^{N} 2
    \langle i |T_m({\mathfrak X})T_m({\mathfrak X})| i \rangle
    -  \mu_{0}  \,,
  \end{split}
\end{equation}
reducing the numerical effort by another factor $1/2$.  Note that the
algorithm requires storage only for the sparse matrix ${\mathfrak X}$
and two vectors of the corresponding dimension.

\section{Reduction of the bandwidth of the Anderson matrix}
\label{RedBB}

\begin{figure}%[ht]
  \begin{minipage}{0.55\linewidth}
    %\centering 
    \includegraphics[width=0.95\linewidth]{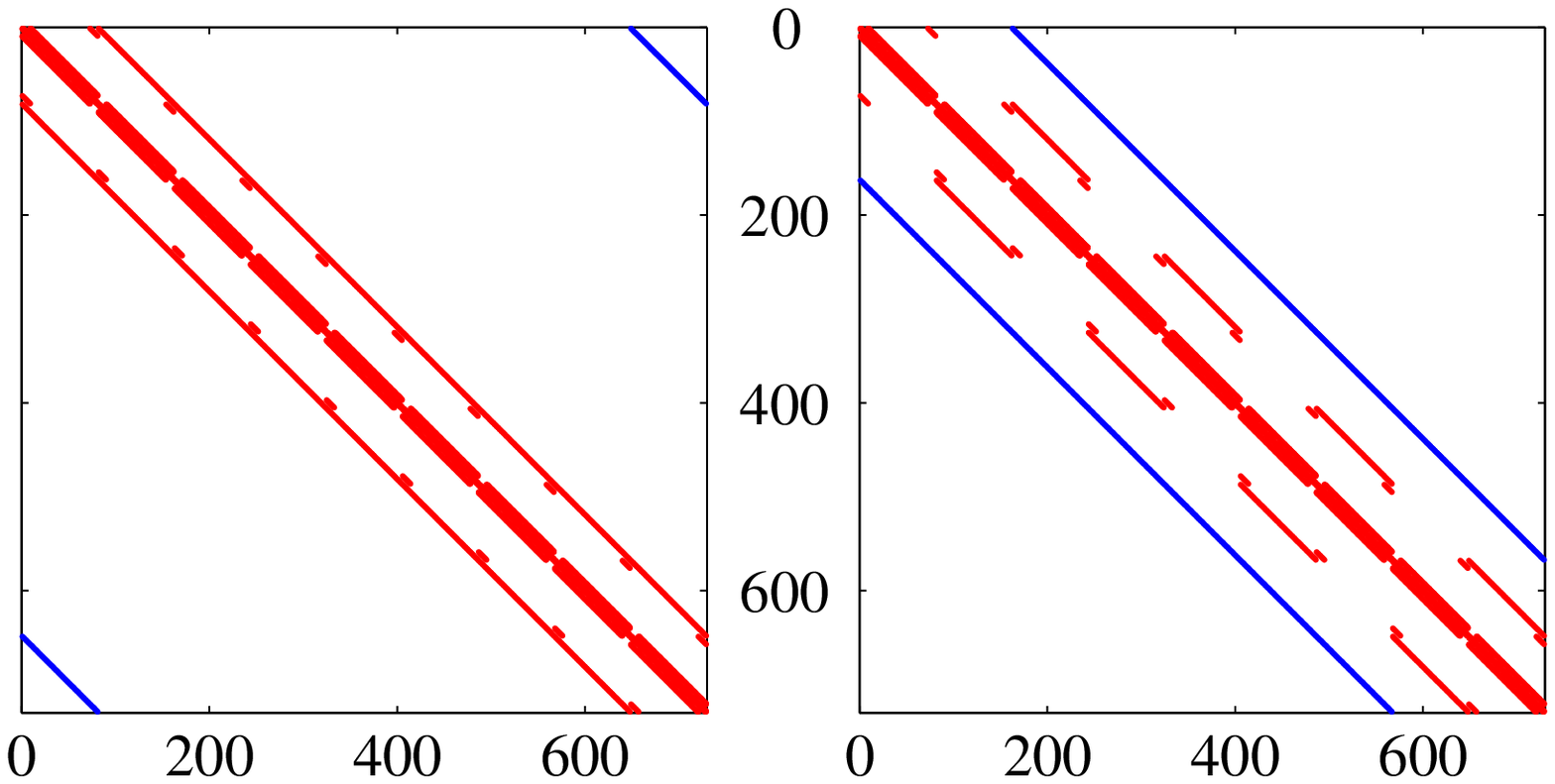}
  \end{minipage}
  \begin{minipage}{0.44\linewidth}
    {\bf Fig.10} 
    Sparsity pattern of the Anderson matrix.
     Left: Standard Anderson matrix for a $9^3$ system with PBC.  The
     use of DSBEV would require the storage of $648$ off-diagonals.
     Right: Reduced matrix after transformation \eqref{RedBBTrafo}.
     Only $162$ off-diagonals have to be stored.
  \end{minipage}
  \label{Red_BB}
\end{figure}
While for Krylov sub-space methods~\cite{Sa92} (like Lanczos or
Jacobi-Davidson) we can take advantage of the sparsity of the
tight-binding type matrices, the full diagonalisation with LAPACK
routines requires their complete storage. Unfortunately, for periodic
boundary conditions the cyclic tridiagonal structure of the
matrices spoils the use of band matrix routines like DSBEV. For a
$L^3$-system there are non-vanishing matrix elements in a distance of
$L^3-L^2$ from the diagonal (Fig.~10).  Thus almost all
matrix elements (most of them zero) need to be stored, giving no
advantage compared to full matrix diagonalisation routines like DSYEV.

For linear systems there are tricks to use tridiagonal matrices
instead of the corresponding cyclic tridiagonal matrices, which are
based on the use of the Sherman-Morrison formula~\cite{PFTV86}. We are
not aware of similar ideas for eigenvalue problems. It turns out,
however, that an appropriate sequence of Givens rotations~\cite{PFTV86}
allows the transformation of the cyclic tight-binding matrix (with
quadratic blocks close to the outer edges) onto a matrix with blocks
only along five diagonals. The corresponding transformation reads
\begin{equation}\label{RedBBTrafo}
    H_{\text{red}} = T^{T}{H}T \, ,
\end{equation}
where $T =  P\otimes{\mathds 1}_{L^2\times L^2}$ and
for odd $L$ the $L\times L$ matrix $P$ is given by
\begin{equation}
  P = \frac{1}{\sqrt{2}}
  \begin{pmatrix}
    \sqrt{2} &  0 & \hdotsfor{6}                    &    0  \\
    0  & -1 & 1 &    0   & \hdotsfor{4}       &    0  \\
    0   & \hdots  & 0 &-1& 1&  0 & \hdotsfor{2} & 0     \\
    \hdotsfor{9}\\
    0  & \hdotsfor{5}           &   0   &   -1   &    1  \\
    0  & \hdotsfor{5}           &   0   &    1   &    1  \\
    \hdotsfor{9}\\
    0  & \hdots  & 0 & 1& 1&  0 & \hdotsfor{2} & 0     \\
    0  & 1 & 1 &    0   & \hdotsfor{4}       &    0  \\
  \end{pmatrix}\,.
\end{equation}
For even $L$ the first row and column are absent. 

The bandwidth of $H$ can thus be substantially reduced (to $2L^2$,
see Fig.~10), which allows for the full diagonalisation of
systems up to $30^3$ sites on PC-systems with a memory of $512$~MB.
Furthermore, the sparsity of the transformation~\eqref{RedBBTrafo} can
be used to avoid an explicit matrix-matrix multiplication~\cite{Sc03}.
Hence the change from $ H$ to $ H_{\text{red}}$ is not time consuming.
The advantage of the transformation is not primarily a gain in CPU
time but storage.

%\bibliography{swf03} 
%\bibliographystyle{apsrev}

\end{document}